# A New Look at an Old Tool – the Cumulative Spectral Power of Fast-Fourier Transform Analysis

Sheng-Chiang Lee[a] and Randall D. Peters

Physics Department, Mercer University, Macon, GA 31207

As an old and widely used tool, it is still possible to find new insights and applications from Fast Fourier Transform (FFT) - based analyses. The FFT is frequently used to generate the Power Spectral Density (PSD) function, by squaring the spectral components that have been corrected for influence from the instrument that generated the data. Although better than a raw-data spectrum, by removing influence of the instrument transfer function, the PSD is still of limited value for time varying signals with noise, due to the very nature of the Fourier transform. The authors present here another way to treat the FFT data, namely the Cumulative Spectral Power (CSP), as a promising means to overcome some of these limitations. As will be seen from the examples provided, the CSP holds promise in a variety of different fields.

---

[a] Electronic Address: lee_sc@mercer.edu

## I. Introduction to the FFT

The development of many Fast Fourier Transform (FFT) algorithms since 1965 has benefited the science and engineering communities in numerous ways. Data acquisition in the time-domain usually contains noise that is either impossible or extremely difficult to suppress. In some cases, the time-domain data even contain too much information to be comprehended in its raw form. Indeed, the authors have frequently encountered difficulty trying to extract important information directly from time-domain data. The solution to this problem is to move interpretation into the frequency-domain where information at certain signature frequencies can be distinguished from the background noises or other unwanted signals. Due to the discrete nature of data acquisition, the Discrete Fourier Transform (DFT) is needed to transform time-domain data into the frequency-domain. The great success of the FFT derives from its greatly improved efficiency—and thus (fast) speed of doing the minimum set of required numerical calculations. The FFT has become so successful that spectral analysis has become commonplace in virtually every research and development field.

## II. Obstacles in FFT applications

In spite of its great power and usefulness, the full potential of the FFT has not been carefully explored. Consider the following example from seismology. In Fig. 1, the FFT was used to operate on conventional time-domain seismograph output to yield the Power Spectral Density (frequency domain) curves shown in **Fig. 1**. Each PSD function was obtained from a 4096-point FFT corresponding to a time-duration of 4096 seconds.

Adjacent FFT's used to generate the curves are separated from one another by a 1024-second time-shift.  To the authors' knowledge, the water-fall display is the best way to show groups of PSD functions of this kind.  Fluctuations in the PSD, due to the very nature of the data and its FFT, inevitably hinder one's ability to study fine-structure features of the spectra that evolve with time.

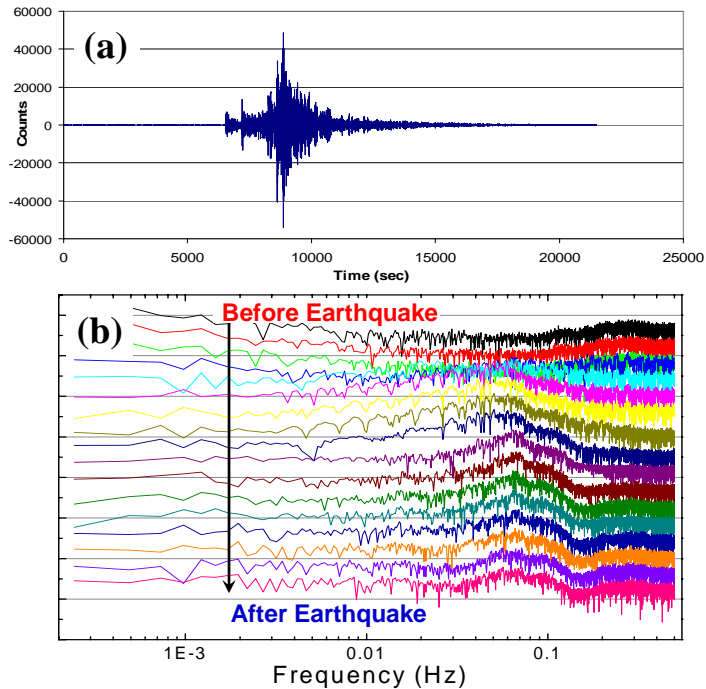

Fig. 1 (a) Record of a teleseismic earthquake in the Kuril Islands, and (b) time-dependent PSD functions from before the arrival of earthquake waves until after they had past.  The instrument of this study is a state-of-the-art digital seismograph called the VolksMeter, described online at http://rllinstruments.com.

## III. The Solution - the Cumulative Spectral Power

To overcome the barrier set by the fluctuations in the PSD, the authors suggest a different approach to frequency domain analyses, namely the Cumulative Spectral Power (CSP).  The CSP is related to the PSD in similar manner as the cumulative probability function (CPF) is related to the probability density function (PDF), as illustrated in Fig. 2.  Both the PDF (a Gaussian distribution with random noises) and the CPF are displayed here. The CPF is calculated with the following equation.

$$F(x) = \int_{-\infty}^{x} f(x')dx' \tag{1}$$

where $F(x)$ and $f(x)$ represent CPF and PDF respectively. The density function f(x') specifies the probability that a variate from the distribution will have a value that lies between x' and x'+ dx'. In turn, F(x) specifies the probability that a given variate will have a value smaller than x.

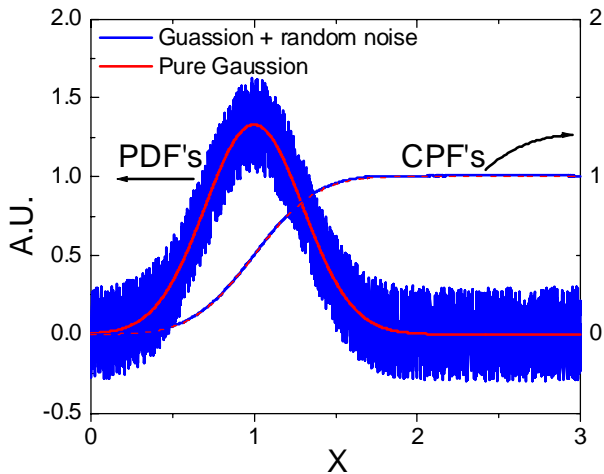

Fig. 2 Comparisons between PDF's and CPF's of the Gaussian distribution with and without significant noise present. Although the noise significantly corrupts the PDF, its CPF shows little difference from the pure Gaussian CPF.

The random noises in the simulated PDF in Fig. 2 have been added to mimic the influence of fluctuations commonly seen in real PSD's obtained from the FFT of time-domain data. Since these fluctuations are mostly random, integration of the PSD reduces their influence while preserving the essential information in a much more perceivable way. As demonstrated in Fig. 2, the 'noises' superposed on the Gaussian distribution are severe. However, it is seen that the influence of that noise is barely perceptible in the CPF that was obtained by integrating the noisy PDF.

## IV. CSP vs. PSD

Since the PSD function is intended to present the power spectral density, as described by the name, the units of the PSD function will naturally be Watt/Hz (kg-m$^2$/s$^3$/Hz) or Watt/kg/Hz (m$^2$/s$^3$/Hz) in the seismometer's case if the mass of the instrument is scaled out to produce what is called the specific power spectral density (SPSD). This is simple and true if the PSD function is plotted with respect to the linear frequency. If it is plotted with respect to the logarithm of the frequency, then the change of variable associated with $df = f \cdot d(\ln(f))$, results in the change of the density function by the factor of frequency, so that the total power of the instrument, which can be obtained by properly integrating the spectra, is invariant under the change of variable.[1]

The CSP presentations, on the other hand, do not suffer the subtlety associated with the change of variable since the CSP simply describes how much power is distributed below (or above) a certain frequency (or period). The units of the CSP are, therefore, m$^2$/s$^3$ if System International units are used to express the CSP.

In studies of real physical signals, e.g. seismic data, the PSD (c.f. Fig. 1) may at its best tell us the frequencies of various waves, but not much else because of the complex nature of the diverse seismic motions that comprise the total spectrum. By contrast, the CSP allows detailed spectral structures to be revealed. For example, the CSP in Fig. 3 clearly shows not only the frequencies of select (well known) seismic waves, but also how energy gets redistributed among various differing frequency regimes as a function of time. Each CSP curve in Fig. 3 was calculated from the corresponding PSD curve in Fig. 1 using Eq. (1), except for the difference that the plot is in terms of period rather than frequency. In the case of the frequency plot, the PSD corresponds to the derivative with respect to frequency of the CSP. In a period plot the PSD corresponds to the negative derivative with respect to frequency of the CSP, specified in terms of period. This subtle difference in the two cases serves to enhance (i) the visibility of low frequency features when plotted vs period and (ii) the visibility of high frequency features when plotted vs frequency.

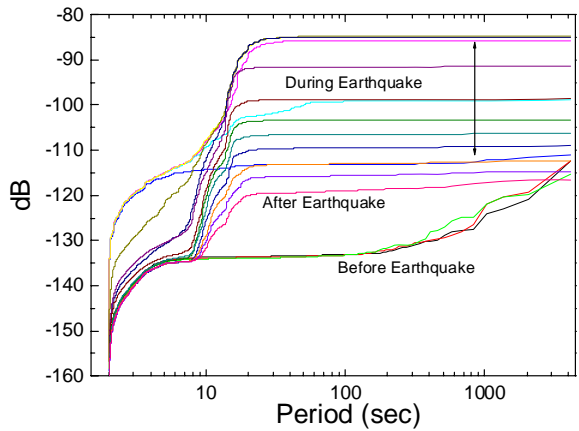

Fig. 3 Time-dependent CSP spectra from before the arrival of earthquake waves until after they had passed. The evolutionary process of redistributing energy over a wide frequency range during this period of time is clearly shown with fine structures indicating some eigen-mode excitations of the earth.

Given the commonly available computational power of personal computers, these analyses can be performed in parallel with the data acquisition process. This allows for near-real-time generation of the CSP curves that are much better than the PSD for observing spectral fine-structures. Fig. 4 shows a screen-shot of the program written by the authors to generate CSP curves in Fig. 3. Only minor modifications are needed to interface the program with real-time data acquisitions from an instrument and generate near-real-time CSP curves.

Another area where the CSP has been shown to reveal much more detailed frequency domain information than is possible with the PSD is that involving material properties. For the study of the evolution of defect structures in solids, real-time CSP analyses open a bright future for understanding internal friction—for which there is still no agreed-upon first principles explanation. It might even happen, and the authors will continue to explore this method, as applied to seismology in an ambitious effort to discover precursors to earthquakes.[2] Additionally, the method is expected to be useful in the study of fractures that occur because of fatigue failure in metals.

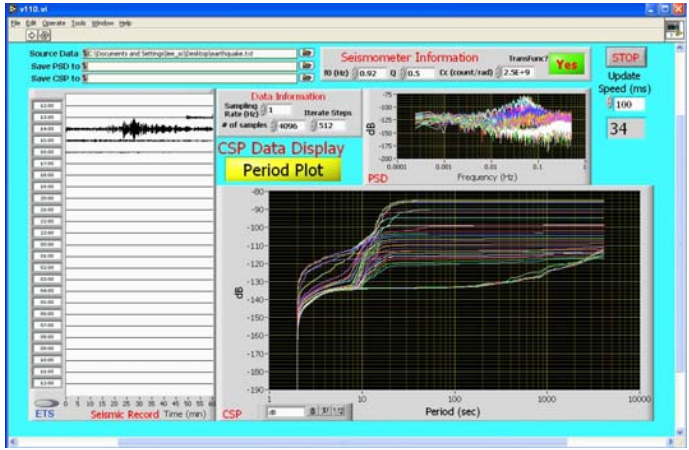

Fig. 4 A screen-shot of the program used to calculate the CSP curves in Fig. 3. The same set of data is used in this screen-shot with a finer time-shift (512 s) between time-windows. All thirty-four CSP curves can be generated in a few seconds.

## V. Potential of CSP analysis to medical applications

Real-time CSP analysis is not limited in benefit to the case so far cited. It is also already influencing cardiovascular diagnosis.[3,4] Traditionally, doctors in cardiology need to have well-trained ears to distinguish various features of abnormal heart beats from normal ones. If an objective diagnostic procedure that does not rely on human hearing ability could be developed, its influence would be far reaching. One of the authors (Peters) has been working in collaboration with the Medical School of Mercer University to assess the value of cardiovascular CSP analysis. Shown in Fig. 5 are CSP records from three different subjects. The two curves labeled as SupraVentricular Tachycardia derive from the same subject, who is capable of transitioning his heart between the two states shown. Additionally, there is a subject whose mitral valve prolapse (heart murmur) shows up clearly by way of the `steps'. And finally, there is the record from the subject without any known heart abnormalities.

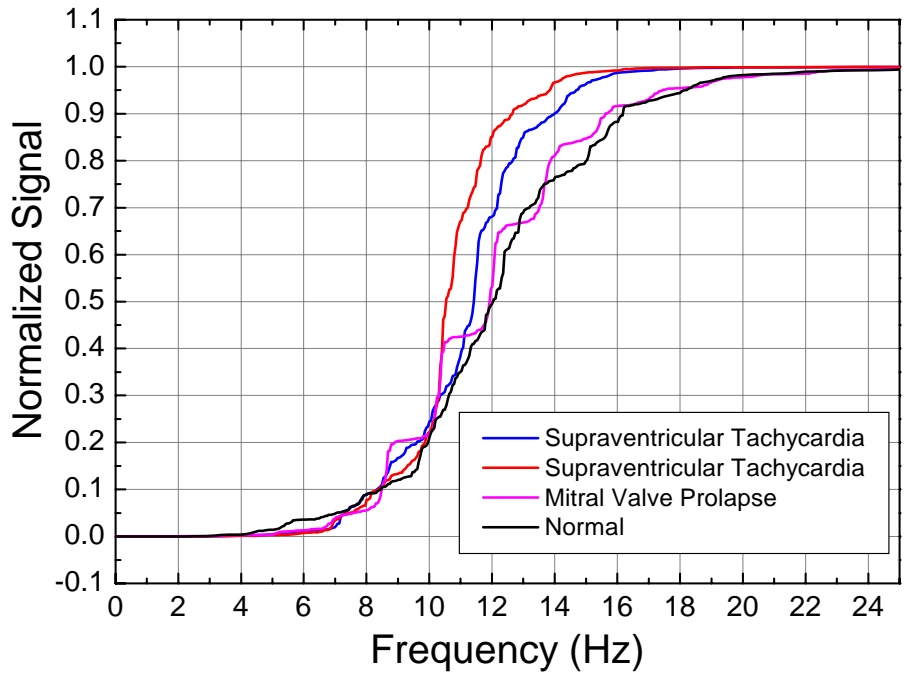

Fig. 5 CSP heart study using a geophone placed on the chest of the subject. Significant differences in energy distribution and modes of excitations are clearly seen among different subjects. This use of the CSP provides substantial improvement in the means to see abnormal heartbeats, as compared to earlier methods of analysis, such as described online at R. D. Peters (2005), "Mechanical cardiography using a geophone", http://physics.mercer.edu/petepag/cardio.html

The CSP records shown in Fig. 5 show clearly distinctive differences. Use of the method of seismocardiography[5] is continuing in the Medical School of Mercer University; additional studies have been planned to involve patients recovering from heart surgery. Provided that the results are as promising as initial cases suggest, the real-time CSP analyses may provide a convenient, objective, accurate, and un-ambiguous way to detect heart diseases at early stages.

## VI. Summary

Although the CSP is simply a new variant of FFT analysis, it provides distinct advantages over other important and well known methods. It is opening new channels for discovery in seismology, materials science concerned with fatigue failure, and even early detection of cardiovascular diseases. Its near-real-time use, because of increasing computational power, has already demonstrated fruitful benefit to the above mentioned fields. No doubt it could also greatly benefit studies in a variety of other fields.

---

[1] R. D. Peters (2007), "A new tool for seismology—the Cumulative Spectral Power", http://arxiv.org/abs/0705.1100

Disagreements from the seismology community in the interpretation of the units of PSD functions, as suggested by numerous personal conversations between one of the authors (RDP) and well-known experts in the field, demand a serious investigation initiated within the scientific/engineering community.

[2] Randall D. Peters, "Precursor Warnings of Structural Catastrophe through observation of a Seismic `Bounce'," http://physics.mercer.edu/hpage/CSP/bounce.html.

[3] I. Perlstein and A. Hoffman, "The cumulative plot of power spectral analysis of heart rate variability assesses the kinetics of action of cholinergic drugs in rats," Engineering in Medicine and Biology Society, 2000. Proceedings of the 22nd Annual International Conference of the IEEE, **1**, pp. 260 - 261, 2000.

[4] K. Howorka, J. Pumprla, P. Haber, J. Koller-Strametz, J. Mondrzyk, and A. Schabmann, " Effects of physical training on heart rate variability in diabetic patients